\documentclass[twocolumn]{aastex}
\usepackage{float}
\usepackage{listings}
\usepackage{amsmath}
\usepackage{graphicx}
\usepackage [english]{babel}
\usepackage [autostyle, english = american]{csquotes}
\MakeOuterQuote{"}

\shortauthors{Munson et al.}

\begin{document}

\title{R Coronae Borealis Star Evolution: Simulating 3D Merger Events to 1D Stellar Evolution Including Large Scale Nucleosynthesis}

\author{Bradley Munson}
\affiliation{Dept. of Physics and Astronomy, Louisiana State University, Baton Rouge, LA 70803, USA}
\author{Emmanouil Chatzopoulos}
\affiliation{Dept. of Physics and Astronomy, Louisiana State University, Baton Rouge, LA 70803, USA}
\author{Juhan Frank}
\affiliation{Dept. of Physics and Astronomy, Louisiana State University, Baton Rouge, LA 70803, USA}
\author{Geoffrey C. Clayton}
\affiliation{Dept. of Physics and Astronomy, Louisiana State University, Baton Rouge, LA 70803, USA}
\author{Courtney L. Crawford}
\affiliation{Dept. of Physics and Astronomy, Louisiana State University, Baton Rouge, LA 70803, USA}
\author{Pavel A. Denissenkov}
\affiliation{Dept. of Physics and Astronomy, University of Victoria, Victoria, BC V8P5C2, Canada}
\author{Falk Herwig}
\affiliation{Dept. of Physics and Astronomy, University of Victoria, Victoria, BC V8P5C2, Canada}

\begin{abstract}
R Coronae Borealis (RCB) stars are rare hydrogen-deficient carbon-rich variable supergiants thought to be the result of dynamically unstable white dwarf mergers. We attempt to model RCBs through all the relevant timescales by simulating a merger event in \textit{Octo-tiger}, a 3D adaptive mesh refinement (AMR) hydrodynamics code and mapping the post-merger object into \textit{MESA}, a 1D stellar evolution code. We then post-process the nucleosynthesis on a much larger nuclear reaction network to study the enhancement of s-process elements. We present models that match observations or previous studies in most surface abundances, isotopic ratios, early evolution and lifetimes. We also observe similar mixing behavior as previous modeling attempts which result in the partial He-burning products visible on the surface in observations. However, we do note that our sub-solar models lack any enhancement in s-process elements, which we attribute to a lack of hydrogen in the envelope. We also find that the \textsuperscript{16}O/\textsuperscript{18}O isotopic ratio is very sensitive to initial hydrogen abundance and increases outside of the acceptable range with a hydrogen mass fraction greater than 10\textsuperscript{-4}.
\end{abstract}


\section{Introduction}
R Coronae Borealis (RCB) stars are rare hydrogen-deficient, carbon-rich supergiants \citep{Clayton12, Clayton96}. They are best known for their irregular variability of up to 8 magnitudes induced by dust formation in their atmospheres. RCB stars are suspected to be a result of a binary white dwarf (WD) merger event consisting of a carbon-oxygen (CO) and a helium (He) WD \citep{Webbink84}. The surface abundances of RCB stars are not only extremely helium-rich and hydrogen-deficient, but also contain an enrichment in N, Al, Na, Si, S, Ni, and several s-process elements \citep{Asplund00}. Furthermore, RCB stars show extraordinarily low \textsuperscript{16}O/\textsuperscript{18}O (on the order of unity) and large \textsuperscript{12}C/\textsuperscript{13}C, which, along with the enrichment of s-process elements, are consistent with partial He-burning \citep{Clayton07}. The mass ratio of the WD merger should be between 0.5 and 0.6 with the total mass ranging between 0.7 and $1.0M_\odot$ based on the requirement of dynamically unstable mass transfer \citep{Dan11, Zhang14} and the high temperature region required for partial He-burning \citep{Staff12}.

Previous attempts to model RCB stars utilize "stellar engineering" (described in Section \ref{sec:comp_structure}) \citep{Lauer19, Crawford20}, modified evolution of zero-age main sequence stars \citep{Weiss87,Menon13,Menon19}, and mapping 3D merger simulations into a 1D stellar evolution program \citep{Schwab19}. Of those methods, the first two attempt to mimic the thermal properties of the post-merger object based on hydrodynamics simulations. This work attempts to model RCBs with the third method, which directly takes into account the thermal structure from the 3D merger model resulting from the dynamical merger.

In this paper, we construct the binary system in a 3D adaptive mesh refinement (AMR) hydrodynamics code called \textit{Octo-tiger} \citep{Marcello16}. After the merger event, we perform a volume weighted spherical average of the evolved parameters. Using Modules for Experiments in Stellar Astrophysics (MESA) version r12115 \citep{Paxton2011,Paxton2013,Paxton2015,Paxton2018,Paxton2019}, we implement the built-in relaxation algorithms to construct a 1D model of the post-merger object and let it evolve through the RCB phase and back down the WD cooling track. Once the stellar evolution calculations are finished, the model is then post-processed with a more complete nuclear network using the \textit{NuGrid} multi-zone nucleosynthesis code \textit{MPPNP} \citep{Herwig08}. This post-processing analysis of the nucleosynthesis happens for every zone at every time step of the stellar evolution and includes 1093 isotopes with their corresponding nuclear reactions.

The models presented in this paper make improvements on what has been done in previous models using smooth particle hydrodynamics (SPH) codes or "stellar engineering" in \textit{MESA} \citep{Dan11,Longland11,Staff12,Menon13,Zhang14,Staff18,Menon19,Lauer19,Schwab19,Crawford20}. In the following sections we compare our initialization process to other works that simulate RCB stars (Section \ref{sec:compare}), present our numerical methods in generating these models (Section \ref{sec:methods}), present the results of our parameters studies on our models (Section \ref{sec:results}), and follow up with concluding thoughts on improvements we will make to these models (Section \ref{sec:conclusion}).

\section{Comparison with Previous Work}
\label{sec:compare}
In this section, we describe how our models are initialized differently than previous studies. In particular, we focus on the thermal structure, the nuclear network, and the mechanisms for chemical mixing.

\subsection{Initialization of Thermal Structure}
\label{sec:comp_structure}
In general, there have been three methods used to generate the thermal structure of an RCB star. Those methods are the He star model, stellar engineering, and mapping a spherically averaged structure from a 3D hydrodynamics code.

The He star model is employed in works such as \cite{Weiss87} and \cite{Menon13, Menon19}. This model is initialized with a helium zero-age main-sequence star and after the envelope has shrunk to some predetermined mass, a composition motivated from realistic progenitor models is imposed on the envelope. \cite{Menon13} splits the abundance profile into four zones which are initialized based on progenitor models. Those zones are the core, a buffer region to prevent dredge up from the core, the He-burning region, and the envelope. The separate initialization of the He-burning region based on nuclear burning during the dynamical phase is a more sophisticated approach that is not considered in our models. This provides a CO core consistent with core He-burning underneath a He-burning shell and an envelope with the expected composition. However, the thermal state of the core is not in agreement with what would be realized in a WD merger scenario and this will have effects on the evolution and state of the envelope \citep{Iben90} (Further discussed in Section \ref{sec:early_ev}). In the WD merger scenario, one should expect the CO WD that would form the core of the RCB to be cooler and more degenerate than the CO core that would form from core He-burning. The additional entropy taken into account at the core-envelope boundary from the dynamical merger event leads to convective mixing in the envelope. \cite{Menon13} do not see a convective envelope and instead require a secular mixing model be adopted to account for the required envelope mixing and justify that the additional mixing may be induced by rotation.

Stellar engineering is a general term used to describe any ad-hoc change to the numerical model that is not implemented through a physical process in order to construct a stellar model that matches some expected initial structure. One way to use stellar engineering to create RCB stars is discussed in great detail in \cite{Lauer19} or \cite{Crawford20}. This process provides a more physically motivated structure than the He star method as it creates a cool and degenerate CO core. In order to get He shell burning and the desired post-merger radius, a process of adjusting the entropy as described in \cite{Shen12} is employed to add entropy to the envelope until the desired post-merger radius of $0.1 R_\odot$ is reached. This method also generates enough mixing via convection to bring species from the He-burning region to the surface.

Mapping data from a 3D hydrodynamics code is the method used in this paper as well as studies such as \cite{Longland11} and \cite{Schwab19}. \cite{Schwab12} took 3D models from \cite{Dan11} and performed 2D simulations of the viscous phase of the merger with $\alpha$-viscosity prescription. Those models evolve into spherical states on the order of a few hours and are then mapped into the 1D \textit{MESA} grid and explored further in \cite{Schwab19}. Our models do not contain a prescription for viscosity and therefore will not naturally evolve into spherical states. Furthermore, there is difficulty when mapping the thermal structure because of any mismatch in the Equation of State (EoS) between two different codes. Like \cite{Schwab19}, our models use the Helmholtz EoS to compute the temperature, which assumes full ionization, and therefore the Helmholtz EoS and \textit{MESA} should match well near the peak temperature region. There is, however, a discrepancy due to the fact that \textit{MESA} assumes the object is in Hydrostatic Equilibrium (HSE), which is not the case for the post-merger object coming out of \textit{Octo-tiger}. More details regarding this discrepancy can be found in Section \ref{sec:Mapping}.

\subsection{Nucleosynthesis and Mixing}
When comparing surface abundances in constructed models to observations, it is important to consider key isotopes and reactions as well as mixing. Most studies include at least a basic nuclear network using H, He, C, N, O, and F, but for investigating s-process nucleosynthesis it is also important to include neutron-source and -capture reactions as well as more massive species. In our models we use the MESA nuclear network \texttt{mesa\_75.net} (herein referred to at \texttt{mesa\_75}) for co-processing as it includes neutrons and species up to \textsuperscript{60}Zn. Post-processing is then done with 1093 species in \textit{MPPNP} and, because of our large co-processed network, our models are usually consistent for the most prominent species. The \textit{MESA} calculated convective regions are used to inform mixing in the post-processing stage as well. While some of the below studies have considered the effects of rotation-driven mixing, we do not utilize those mixing prescriptions because convective mixing was sufficient in bringing the partial He burning products to the surface.

\cite{Lauer19} used \texttt{mesa\_75} for co-processing as well, but does not perform any post-processing. However, they see good agreement with \textsuperscript{16}O/\textsuperscript{18}O and C/O as well as surface abundances during the RCB phase. In terms of mixing, both convection and rotationally induced mixing are used in order to transport the species from the burning region to the surface. However, only solar metallicities are considered in their study, but the observations of most RCB stars show sub-solar metallicities \citep{Asplund00,Clayton07}. \cite{Crawford20} expands upon those models by analyzing models with sub-solar metallicities using the same stellar engineering method.

\cite{Menon13,Menon19} study solar and sub-solar metallicities, respectively. These studies take extra care to analyze the nucleosynthesis responsible for the overabundance in \textsuperscript{18}O and \textsuperscript{19}F. They find that they are only able to reproduce the surface abundances with a particular mixing profile that mixes material not so fast as to destroy all of the \textsuperscript{18}O but fast enough to bring \textsuperscript{15}N to the burning region in order to form \textsuperscript{19}F. This mixing profile is implemented by creating an additional diffusion coefficient which is justified to be the result of rotationally induced mixing. Furthermore, this additional diffusion coefficient must halt before the RCB phase in order to maintain the surface \textsuperscript{14}N abundance. Their study also uses post-processing with over 1000 isotopes and investigates s-process elements but finds only lower mass ratios (q=0.5) have a neutron number density high enough to produce s-process elements.

\cite{Zhang14} build a custom nuclear network within \textit{MESA} containing 35 species up to \textsuperscript{32}S and 113 reactions. However, this network does not include neutrons and thus cannot produce s-process elements. As is the case in our study and other studies, mixing via the Mixing Length Theory (MLT) prescription in \textit{MESA} is sufficient to bring species from the burning region to the surface \cite{Lauer19,Crawford20}. 

\cite{Schwab19} chooses not to focus on the nucleosynthesis given the limitations of the nuclear network used in \cite{Schwab12}. However, they do emphasize the importance of a modified opacity table as the default opacity tables used in \textit{MESA} are not well suited for following the evolutionary tracks of RCB stars. The default opacity table in \textit{MESA} is calculated using GS98 \citep{gs98} solar scaled abundances. Although \textit{MESA} does include opacity tables for carbon and oxygen-rich mixtures, referred to as "Type 2" tables, the lower temperature boundary for those tables is $\log{(T/K)} = 3.75$. This will not be consistent with RCB stars that develop cooler envelopes because the outer layers are hydrogen deficient and carbon and oxygen enhanced. This has an effect on the effective temperature and radius during the RCB phase, but does not appear to have a noticeable effect on surface composition. Therefore, we only use the default \textit{MESA} "Type 2" opacity tables and do not explore this parameter further.

\section{Methods}
\label{sec:methods}
Our post-merger simulations build on previous work done using "stellar engineering" in \textit{MESA} \citep{Lauer19}. In the previous approach, a post-merger structure was constructed in order to mimic a thermal and chemical structure consistent with that found in past 3D hydrodynamics merger simulations. The effects of rotation were included by assuming solid body rotation with 20\% break-up velocity. Then, the post-merger evolution is followed to the RCB phase using the MESA 75-isotope nuclear network, \texttt{mesa\_75}. The surface abundances, surface rotation rates, and the time spent in different phases were computed for several models with differing mass parameters.

The work presented here uses one 3D hydrodynamics model that has either a solar or sub-solar metallicity composition imposed onto the post-merger object. In each choice of metallicity, the evolution of the post-merger object is followed through the RCB phase and back to the WD state. The initial conditions and results of each model are summarized in Table \ref{tab:1}. The following subsections outline the procedure used to map the 3D hydrodynamics grid into \textit{MESA} and process the RCB data. 

\subsection{Progenitor Evolution}
The first step in RCB evolution is a CO+He WD merger event. In order to form a single star, there are two important requirements for the close-binary WD \citep{Zhang14}. First, the mass parameters and the initial separation must be such that the system loses angular momentum and causes the separation to decay until the He WD fills its Roche lobe. Second, once the He WD begins to lose mass, its radius will increase more rapidly than the separation due to the transfer of angular momentum. These two requirements will lead to a dynamically unstable merger event and constrain the mass ratio between 2/3 and 1. However, there are situations in which the spin-orbit coupling results in a dynamically unstable merger event while having a mass ratio less than 2/3 \citep{Dan11}. These constraints are illustrated in Figure 1 of \cite{Zhang14}.

Since the surface abundances of RCB stars are consistent with partial He-burning, it is important to have a hot shell above the CO core that has a sufficient temperature to ignite He-burning. \cite{Staff12} investigate the effects of the binary mass ratio on He-burning shell temperature and find that the temperature decreases as the mass ratio increases. Of course, the He-burning temperature will also increase as the total mass of the system increases. The necessity for a He-burning shell and an unstable dynamical merger constrain the total mass and mass ratio to 0.7-$1.0M_\odot$ and 0.5-0.7, respectively.

Our 3D hydrodynamics simulation, \textit{Octo-tiger}, is initialized with a $0.53M_{\odot}$ CO WD and a $0.32M_{\odot}$ He WD. \textit{Octo-tiger} uses a simple zero-temperature WD (ZTWD) and ideal gas EoS. The ZTWD EoS assumes a zero temperature electron gas with a mean molecular weight per free electron of 2 atomic units. The mean molecular weight for ions and electrons used in the ideal gas EoS is calculated for fully ionized He for the secondary (4/3 atomic units) and an equal mixture of fully ionized C and O by mass for the primary (1.75 atomic units). The pressure and energy equations used for the ZTWD EoS are shown in \cite{Benz90} and are shown below in a slightly different form.

\begin{multline}
    P_{deg} = A\biggl[\left(x(2x^2-3)(x^2+1)^\frac{1}{2}+3\sinh^{-1}{x}\right)\biggr]
\end{multline}

\begin{multline}
    E_{deg} = A\biggl[8x^3\left((x^2+1)^\frac{1}{2}-1\right)\\-\left(x(2x^2-3)(x^2+1)^\frac{1}{2}+3\sinh^{-1}{x}\right)\biggr]
\end{multline}

Where

\begin{equation}
    x = \left(\frac{\rho}{B}\right)^{\frac{1}{3}}
\end{equation}

\begin{equation}
    \begin{aligned}
        A &= \frac{\pi m_e^4c^5}{3h^3}, \\
        B &= \frac{8\pi m_p \mu_e}{3}\left(\frac{m_ec}{h}\right)^3
    \end{aligned}
\end{equation}

\begin{figure}[htb!]
    \centering
    \includegraphics[width=0.99\linewidth]{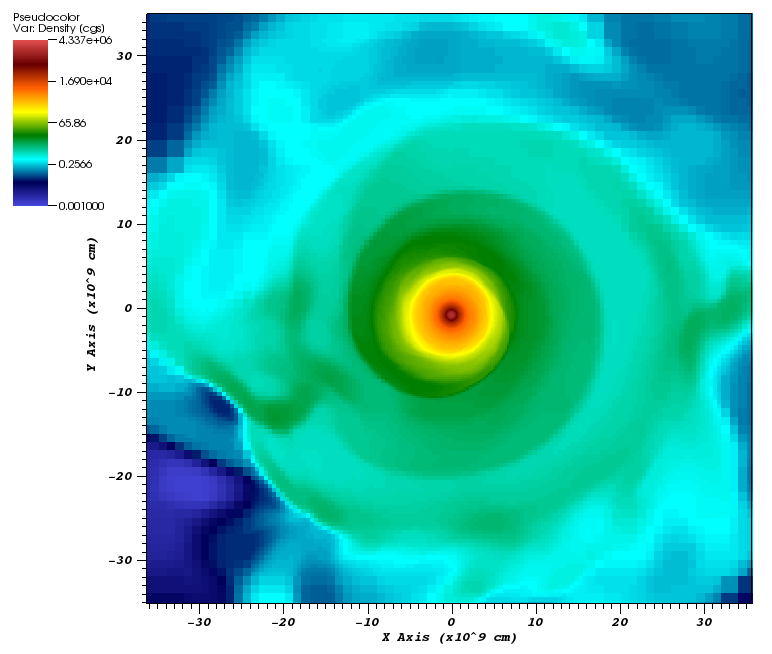}
    \caption{Pseudocolor plot of the density in a slice of the equatorial plane after 3.3 hours of evolution time. This is from the simulation of a $0.53M_\odot$ CO + $0.32M_\odot$ He WD merger.}
    \label{fig:densz}
\end{figure}

\begin{figure}[htb!]
    \centering
    \includegraphics[width=0.99\linewidth]{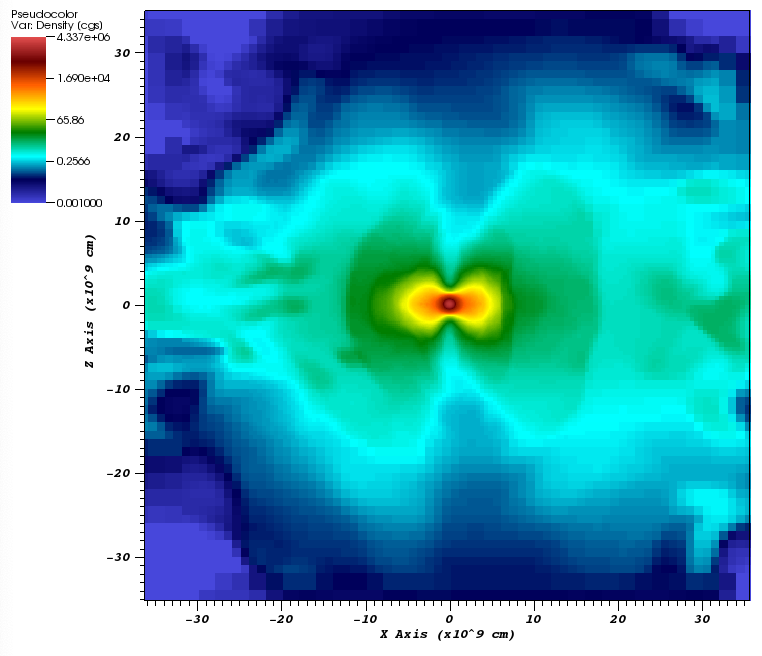}
    \caption{Pseudocolor plot of the density in a slice of the polar plane after 3.3 hours of evolution time. Details of the merger are described in Figure \ref{fig:densz}.}
    \label{fig:densy}
\end{figure}

\begin{figure}[htb!]
    \centering
    \includegraphics[width=0.99\linewidth]{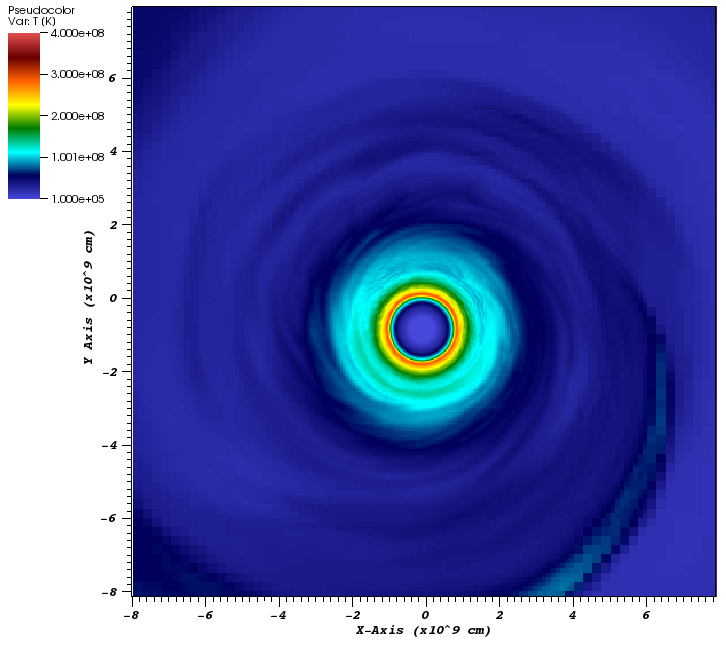}
    \caption{Pseudocolor plot of the temperature in a slice of the equatorial plane after 3.3 hours of evolution time. The He-burning region is apparent around the CO core. Since temperature is not an evolved variable in \textit{Octo-tiger}, a simple calculation using the internal energy density, mass density, and the mean molecular weight of fully ionized helium (4/3) is done for the purposes of this illustration. A more rigorous calculation is done for the spherically averaged model in Section \ref{sec:Mapping}. Details of the merger are described in Figure \ref{fig:densz}.}
    \label{fig:tempz}
\end{figure}

\begin{figure}[htb!]
    \centering
    \includegraphics[width=0.99\linewidth]{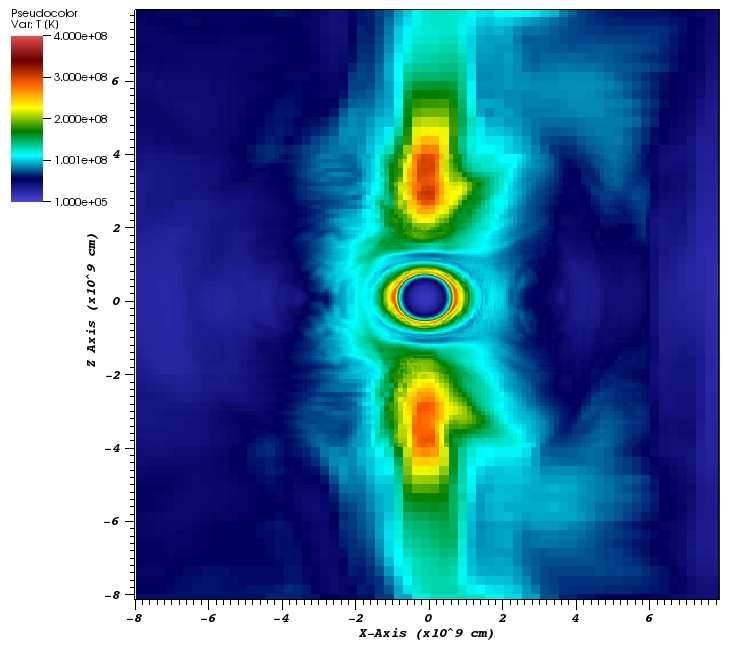}
    \caption{Pseudocolor plot of the temperature in a slice of the polar plane after 3.3 hours of evolution time. The He-burning region is apparent around the CO core. Details of the merger are described in Figure \ref{fig:densz}.}
    \label{fig:tempy}
\end{figure}

While the primary and secondary stars are tracked as fluids, \textit{Octo-tiger} does not trace abundances and therefore we cannot directly obtain a composition profile from this simulation. More details of the composition are discussed in Section \ref{sec:Mapping}. The initial orbital period of the binary is 149 seconds and the initial grid is 24,000 sub-grids with up to seven levels of refinement. The refinement criterion is based solely on the density. Because \textit{Octo-tiger} does not handle gravitational radiation nor magnetic fields, angular momentum is removed from the system at a user-defined rate. For this system, the dynamically unstable merger event initiates after about an hour. After 3.3 hours of evolution in \textit{Octo-tiger}, we spherically average the grid to be mapped into \textit{MESA}. Equatorial and polar slices of the density and temperature are shown in Figures \ref{fig:densz}, \ref{fig:densy}, \ref{fig:tempz}, and \ref{fig:tempy}. In Figures \ref{fig:tempz} and \ref{fig:tempy}, a hot shell is clearly visible and has a peak temperature of about 285 MK, which is notably lower than the models in \cite{Lauer19}, but still high enough to initiate He-burning as seen in Figure \ref{fig:init_base}.

In order to compute the spherically averaged parameters, the following procedure was implemented. First, the center of mass was found by finding the 3D grid cell with maximum density, then all positions are translated so that the center of mass is at the origin. Next, we create roughly 350 spherical shells centered on the origin with radius $r$ and thickness $dr$ such that $\log{\frac{dr}{r}}$ is constant. This binning in log-space and resolution helps to ensure that we get good resolution near the He-burning region and in the outer layers of the star. For each shell, we sum the product of a parameter $X_i$ and cell volume $dV_i$ for all cells in between a radius $r$ and $r+dr$ and divide by the volume of the shell to get the spherically averaged parameter $X_r$ at radius $r$ (shown in Equation \ref{eq:average}). Additionally, we compute the total energy of each shell as the sum of the spherically averaged bulk kinetic energy and a calculated potential energy (Equation \ref{eq:pot_energy}). The radius of the spherical post-merger object is then taken to be the outer radius where the material is bound. For this system, we obtain a radius of $0.50 R_\odot$ and a total bound mass of $0.80 M_\odot$. We note that the difference between the bound mass and the initial masses of the progenitors indicates a mass loss of $0.05M_\odot$, which is likely overestimated by the spherical averaging procedure. 

\begin{equation} \label{eq:average}
    X_r = \frac{\sum X_idV_i}{\sum dV_i}
\end{equation}

\begin{equation} \label{eq:pot_energy}
    \Phi(r) = -\frac{GM_r}{r} - 4\pi G \int_{r}^{\infty} r' \rho (r') dr'
\end{equation}

One concern with taking the spherical average at this point in the hydrodynamics simulation is that the post-merger object is clearly not spherical in Figure \ref{fig:densy}. However, \cite{Schwab12} find that these systems tend towards spherical states on the time-scale of a few hours. We attribute the toroidal shape of our system to the fact that \textit{Octo-tiger} does not have a prescription for viscosity and thus, other than numerical diffusion, there is no physical mechanism to diffuse angular momentum. We find that spherically averaging this system reduces the He-burning region temperature by a factor of two, which is likely due to the non-spherical state of the system "smearing" the peak temperature over the polar angle. It is important to note here that due to the high sensitivity that nuclear reactions have on temperature, these small factors could result in large changes in composition during post-merger evolution. 


\subsection{Mapping into the Spherical Grid of MESA}
\label{sec:Mapping}
The second step of the RCB simulation is the nuclear timescale evolution from the post-merger object through the RCB phase back to a WD state. This evolution is done by mapping our spherically averaged post-merger object into \textit{MESA} using the \textit{MESA} built-in relaxation algorithms. A discussion of how \textit{MESA} relaxes models can be found in Section B of the Appendix in \cite{Paxton2018}. In order to map this object into \textit{MESA}, three profiles are required: composition, specific angular momentum, and entropy. Assumptions and methods used to obtain those profiles are outlined in the remainder of this subsection.

Assuming both components of the post-merger object are well mixed, we impose two different uniform compositions for the CO- and He-WD. The composition is obtained by running the \textit{MESA} test suite problems \texttt{make\_co\_wd} and \texttt{make\_he\_wd} for the solar and sub-solar metallicities. The metallicity fractions used in these models were the \textit{MESA} default options from \cite{gs98}. We then impose the CO WD composition on the post-merger object up to the mass coordinate corresponding to the original mass of the CO WD ($0.53M_\odot$ in our system). Next, we impose the He WD composition from the mass coordinate of CO WD to the surface. An illustration of this composition can be found for some key elements in Figure \ref{fig:init_comp}. This assumes that any mass lost in the system is only lost by the He WD, which is a good assumption since the He WD experienced a tidal disruption while the CO WD remained mostly inert and intact. Because \textit{Octo-tiger} does track the components as fluids, one could in principle use the component fractions in each cell to calculate an abundance assuming the WDs were well mixed and had a homogeneous composition. However, these AMR simulations tend to overestimate the amount of material dredged up from the primary compared to other hydro simulations \citep{Staff12,Staff18}. \cite{Staff18} claim that mixing during the dynamical merger phase has important effects on the surface abundance. In particular, it is important to avoid dredging up too much material from the primary as it could create an overabundance of \textsuperscript{16}O and make it difficult to produce a sufficient amount of \textsuperscript{18}O in order to obtain the isotopic ratio of order unity. Therefore, we find the method of imposing a composition based on mass coordinate to be more reliable.

The angular momentum comes directly from the 3D hydrodynamics grid by computing the axially symmetric angular momentum from the linear momentum and the radius as seen in Equation \ref{eq:am}. After computing the specific angular momentum for each cell, it is averaged using Equation \ref{eq:average} with cylindrical shells.

\begin{equation} \label{eq:am}
    \Vec{j} = \Vec{r} \times \Vec{p} = x p_y - y p_x
\end{equation}

The immediate problem with obtaining the entropy profile from \textit{Octo-tiger} is that the entropy is not an evolved variable and uses a much simpler EoS than \textit{MESA}. The zero temperature WD and ideal gas EoS of \textit{Octo-tiger} will not match the EoS tables used by \textit{MESA}, especially in extreme regions of high density and high temperature. Figure 50 of \cite{Paxton2019} shows the density-temperature coverage of the EoS used in \textit{MESA}. Rather than directly relax an entropy profile, \textit{MESA} can compute an entropy profile by relaxing a density and temperature profile. This is the method we implement because it allows us to compare density and temperature profiles to the previous work in \cite{Lauer19} and we can produce an informed temperature structure of the CO core. The density profile is simply spherically averaged from \textit{Octo-tiger}, but the temperature profile needs to be computed using the evolved internal energy density. Using the \textit{Octo-tiger} internal energy density and the composition profile we imposed on the post-merger object, we compute a temperature profile using the Helmholtz EoS\footnote{http://cococubed.asu.edu/code\_pages/eos.shtml} \citep{Timmes00}. Because \textit{Octo-tiger} assumes a zero temperature WD EoS in the highly dense CO core, we floor this temperature to a constant value of 10 MK. The density and temperature profiles are then combined for relaxation in \textit{MESA}.

\begin{figure*}[htb!]
    \centering
    \includegraphics[width=0.99\linewidth]{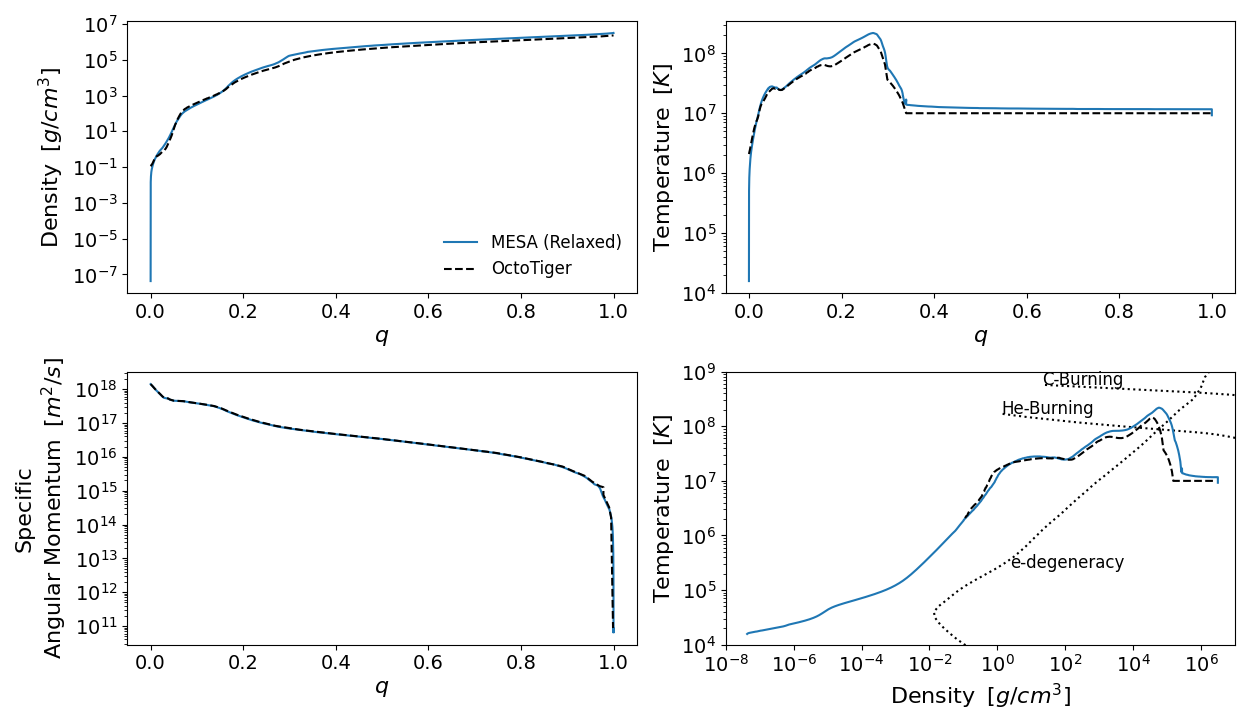}
    \caption{A comparison of the spherically averaged final \textit{Octo-tiger} output (dashed) to the \textit{MESA} relaxed output (solid). $q$ is the normalized exterior mass coordinate ($q=1-\frac{M_r}{M_{tot}}$)}
    \label{fig:init_base}
\end{figure*}

\begin{figure}[htb!]
    \centering
    \includegraphics[width=0.99\linewidth]{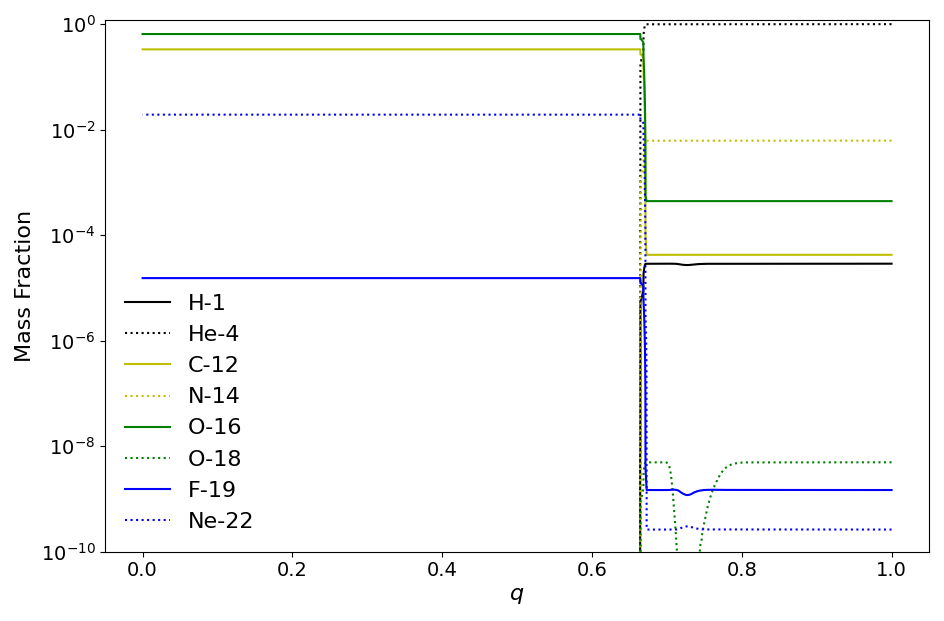}
    \caption{Abundance profiles for key species in Model 1 from Table \ref{tab:1}. The effects of nuclear burning in the hot He shell (or SOF) that has taken place up to that phase can be seen around $q$ $\sim$ 0.7-0.8.}
    \label{fig:init_comp}
\end{figure}

After computing the necessary three profiles, the \textit{MESA} relaxation routine is implemented. Figure \ref{fig:init_base} shows some of the profiles of the spherically averaged \textit{Octo-tiger} output and the relaxed \textit{MESA} model for comparison. Figure \ref{fig:init_base} also shows a density-temperature profile for the \textit{Octo-tiger} output and the \textit{MESA} relaxed model with the He-burning and C-burning regions illustrated. Of these profiles, the only parameter that has a significant difference between the \textit{Octo-tiger} value and \textit{MESA} relaxed value is the temperature (in which the peak relaxed temperature is a factor of 1.5 higher than the peak \textit{Octo-tiger} temperature). We attribute this to the conversion of the \textit{Octo-tiger} internal energy not being hydrodynamically stable with the density profile in \textit{MESA}. However, the peak temperature is still high enough to commence He-burning, a necessary ingredient to obtain the surface abundances observed in RCB stars. It should, however, be noted that even a small range of peak temperature values may significantly change the outcome of the surface abundances during the RCB phase. \cite{Crawford20} analyze the effect of peak temperature on surface abundance during the RCB phase.

\subsection{Nucleosynthesis Post-Processing}
The last step in the RCB simulation is to post-process the data with a much larger scale nuclear network using the \textit{NuGrid} \textit{MPPNP} code \citep{Herwig08}. Since \textit{MPPNP} is a post-processing code, it does not change the evolution of the star, but it performs nucleosynthesis calculations for all isotopes - including those involved in the s-process. This builds in the assumption that the energy produced from the high mass nucleosynthesis does not contribute to the stellar structure and the \texttt{mesa\_75} nuclear network accounts for almost all of the nuclear energy generation. The post-processing code also mixes the species after each time step in accordance with the \textit{MESA} simulation.

After evolving the post-merger object, we take the stellar profile at the point where the surface temperature is at a minimum and luminosity is at a maximum; this is how we define the RCB phase. At this phase we calculate the surface abundances as well as isotopic ratios of interest for RCB stars. Table \ref{tab:1} contains the isotopic ratios and Figure \ref{fig:surf_abund} contains the surface abundances at the RCB phase for the best matched models. The surface abundances are calculated using the typical Equation \ref{eq:surf}. In this equation, $X$ is the average mass fraction of an element of all zones with optical depth less than 1. $\mu_X$ is the mean atomic mass of that element and $\log{\epsilon(X)_\odot}$ is the solar value of that element taken from \cite{Lodders03}. This value represents the log of the number of ions of element $X$ in a sample containing 10\textsuperscript{12.15} ions normalized to the solar value.

\begin{equation}
\label{eq:surf}
[X] = \log{X} - \log{\mu_X} + 12.15 - \log{\epsilon(X)_\odot}
\end{equation}

\begin{table*}[htb!]
  \centering
  \begin{tabular}{ccccccccc}
    \hline
    \hline
 Model & $M_{tot} (M_{\odot})$ & $M_{CO} (M_{\odot})$ & Metallicity & Nuclear Net & \texttt{overshoot\_f} & Initial $^{1}$H & $^{16}$O/$^{18}$O & C/O \\
 \hline
1     & 0.8  & 0.53  & Solar       & MESA75      & 0            & $2.86\times10^{-5}$   & 3.05    & 75.56  \\
2     & 0.8  & 0.53  & Solar       & MESA75      & 0            & $10^{-99}$   & 1.96    & 64.87  \\
3     & 0.8  & 0.53  & Solar       & MESA75      & 0            & $10^{-20}$   & 1.17    & 52.36  \\
4     & 0.8  & 0.53  & Solar       & MESA75      & 0            & $10^{-10}$   & 1.62    & 59.81  \\
5     & 0.8  & 0.53  & Solar       & MESA75      & 0            & $10^{-6}$   & 2.38    & 69.33  \\
6     & 0.8  & 0.53  & Solar       & MESA75      & 0            & $10^{-4}$   & 3.79    & 68.56  \\
7     & 0.8  & 0.53  & Solar       & MESA75      & 0            & $10^{-3}$   & 158.63  & 11.71  \\
8     & 0.8  & 0.53  & Solar       & MESA75      & 0.014        & $2.86\times10^{-5}$   & 1.37    & 58.67  \\
9     & 0.8  & 0.53  & Solar       & MESA75      & 0.02         & $2.86\times10^{-5}$   & 0.91    & 45.97  \\
10    & 0.8  & 0.53  & Solar       & MESA75      & 0.03         & $2.86\times10^{-5}$   & 0.73    & 39.75  \\
11    & 0.8  & 0.53  & Solar       & MESA75      & 0.05         & $2.86\times10^{-5}$   & 0.72    & 37.68  \\
12    & 0.8  & 0.53  & Solar       & MESA75      & 0.055        & $2.86\times10^{-5}$   & 0.81    & 40.09  \\
13    & 0.8  & 0.53  & Solar       & MESA75      & 0.06         & $2.86\times10^{-5}$   & 0.84    & 40.67  \\
14    & 0.8  & 0.53  & Solar       & MESA75      & 0.065        & $2.86\times10^{-5}$   & 1.18    & 36.89  \\
15    & 0.8  & 0.53  & Solar       & MESA75      & 0.068        & $2.86\times10^{-5}$   & 2.50    & 25.16  \\
16    & 0.8  & 0.53  & Solar       & MESA75      & 0.07         & $2.86\times10^{-5}$   & 4.17    & 17.39  \\
17    & 0.8  & 0.53  & Solar       & MESA75      & 0.073        & $2.86\times10^{-5}$   & 8.79    & 10.04  \\
18    & 0.8  & 0.53  & Solar       & MESA75      & 0.075        & $2.86\times10^{-5}$  & 11.85   & 5.96   \\
19    & 0.8  & 0.53  & Solar       & MESA75      & 0.1          & $2.86\times10^{-5}$   & 155.09  & 1.33   \\
20    & 0.8  & 0.53  & Solar       & MESA75      & 0.14         & $2.86\times10^{-5}$   & 1444.50 & 4.00   \\
21    & 0.8  & 0.53  & Solar       & MPPNP      & 0.073        & $2.86\times10^{-5}$   & 8.91    & 6.95   \\
22    & 0.8  & 0.53  & Sub-Solar   & MESA75      & 0            & $7.20\times10^{-5}$   & 33.77   & 282.61 \\
23    & 0.8  & 0.53  & Sub-Solar   & MESA75      & 0            & $10^{-99}$   & 4.28    & 330.88 \\
24    & 0.8  & 0.53  & Sub-Solar   & MESA75      & 0            & $10^{-20}$   & 5.07    & 331.87 \\
25    & 0.8  & 0.53  & Sub-Solar   & MESA75      & 0            & $10^{-10}$   & 5.45    & 347.88 \\
26    & 0.8  & 0.53  & Sub-Solar   & MESA75      & 0.06         & $10^{-99}$   & 1.23    & 184.19 \\
27    & 0.8  & 0.53  & Sub-Solar   & MESA75      & 0.065        & $10^{-99}$   & 4.46    & 81.36  \\
28    & 0.8  & 0.53  & Sub-Solar   & MESA75      & 0.068        & $10^{-99}$   & 14.37   & 30.72  \\
29    & 0.8  & 0.53  & Sub-Solar   & MESA75      & 0.07         & $10^{-99}$   & 15.20   & 31.79  \\
30    & 0.8  & 0.53  & Sub-Solar   & MESA75      & 0.073        & $10^{-99}$   & 55.50   & 9.69   \\
31    & 0.8  & 0.53  & Sub-Solar   & MESA75      & 0.068        & $10^{-10}$   & 10.60   & 39.79  \\
32    & 0.8  & 0.53  & Sub-Solar   & MPPNP      & 0.068        & $10^{-10}$   & 12      & 26.91  
  \end{tabular}
  \caption{Initial conditions and isotopic results of all 32 models. "\texttt{overshoot\_f}" is the extension beyond the convective zone boundary in fractional scale heights. Initial \textsuperscript{1}H refers to the initial mass fraction of \textsuperscript{1}H present in the helium envelope.}
  \label{tab:1}
\end{table*}

\section{Results}
\label{sec:results}
In this section, we discuss the results from our evolutionary models. We start by discussing our initial models for solar and sub-solar metallicities. Next, we discuss variations of those cases as we implement varying degrees of overshooting and change the initial abundance of hydrogen in the RCB envelope. Finally, we use the models that best match observations from the solar and sub-solar cases and post-process them with the \textit{NuGrid} \textit{MPPNP} code for analysis of s-process elements.

\subsection{Initial 3D to 1D \textit{MESA} models}
In this section we discuss the results from our base solar and sub-solar models that contain no overshooting and no initial hydrogen adjustment (Model 1 and Model 22) for comparison against the engineered models of \cite{Lauer19}. Specifically, we focus on the early stages of evolution following the merger and the mixing that takes place. Then, we discuss how changes in overshooting or initial hydrogen abundance affect the results.

\subsubsection{Early Stage Evolution}
\label{sec:early_ev}
\begin{figure}[htb!]
    \centering
    \includegraphics[width=0.99\linewidth]{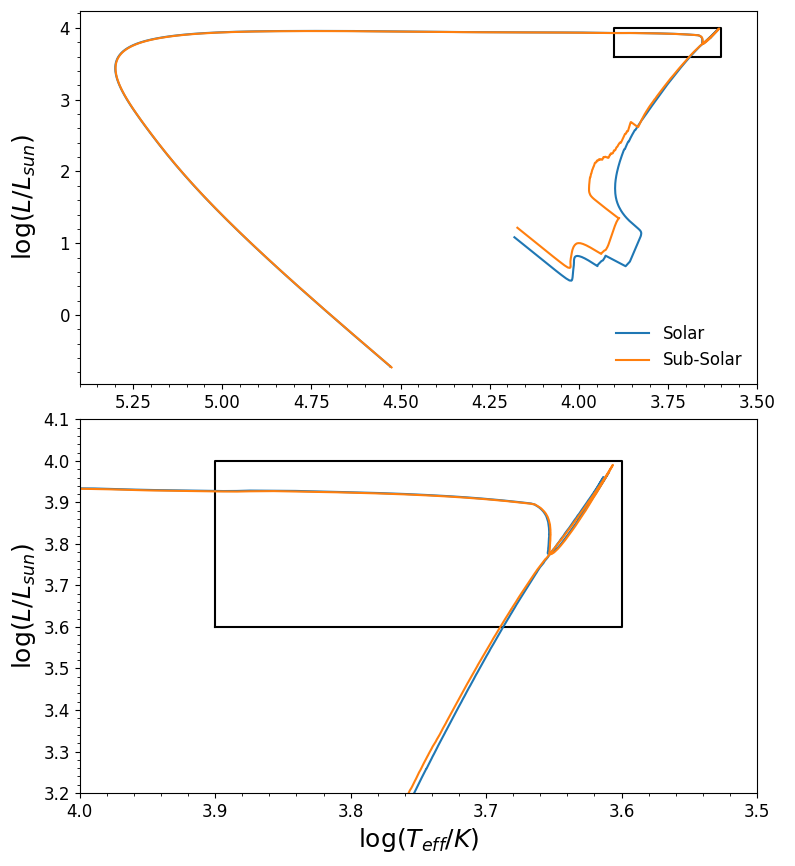}
    \caption{The HRD track for solar and sub-solar metallicity RCB stars (Models 1 and 22 from Table \ref{tab:1}, respectively). The black box represents the range of effective temperatures and luminosities of observed RCB stars.}
    \label{fig:HR}
\end{figure}

The primary goal of this study is to obtain results similar to observations and compare to those obtained by the stellar engineering procedure of \cite{Lauer19}. In their study, they explore a parameter space consisting of initial radius, total mass, mass ratio, rotation, and initial hydrogen abundance. Of those initial parameters, their model A7 is most similar to our initial conditions in terms of total mass, mass ratio, and initial hydrogen ratio. The major differences in our model are that the initial rotational profile, thermal profile, and radius of the post-merger object are calculated based on the \textit{Octo-tiger} grid. After mapping the 3D merger object from \textit{Octo-tiger} into \textit{MESA} using the procedure outlined in Section \ref{sec:methods}, we let it evolve and obtain an HRD track shown in Figure \ref{fig:HR}.

The HRD tracks in this study are similar to those of \cite{Lauer19} and \cite{Schwab19} with an early brightening phase leading the model into the RCB box (the observed range of effective temperatures and luminosities of RCB stars) from the bottom. This is different than the models of \cite{Weiss87} or \cite{Menon13, Menon19} which see a small brightening and cooling phase causing the models to enter the box from the left. This is the result of the 3D merger and engineered models starting off with more compact and cooler cores. The envelope then expands as a result of the energy released from the steady He-burning shell. There is a short period of thermal adjustment in the envelope as the He-burning shell reaches a steady state, after which the solar and sub-solar models have identical tracks in the HRD. This thermal adjustment lasts around 500 years and expands the envelope until $\log(L/L_{\odot})$ reaches 2.5-2.7. \cite{Schwab19} also sees this thermal readjustment phase in their multidimensional model (ZP4) mapped into \textit{MESA} with similar lifetimes (though that model starts off much brighter than ours). The time it takes for the solar and sub-solar models to enter the RCB phase (minimum effective temperature) are 1600 years and 1300 years, respectively. This is in agreement with the lifetimes reported in \cite{Lauer19} and the higher He-burning temperature models of \cite{Crawford20}.

\subsubsection{Surface Abundance and Mixing}
\label{sec:surf_and_mix}
\begin{figure}[htb!]
    \centering
    \includegraphics[width=0.99\linewidth]{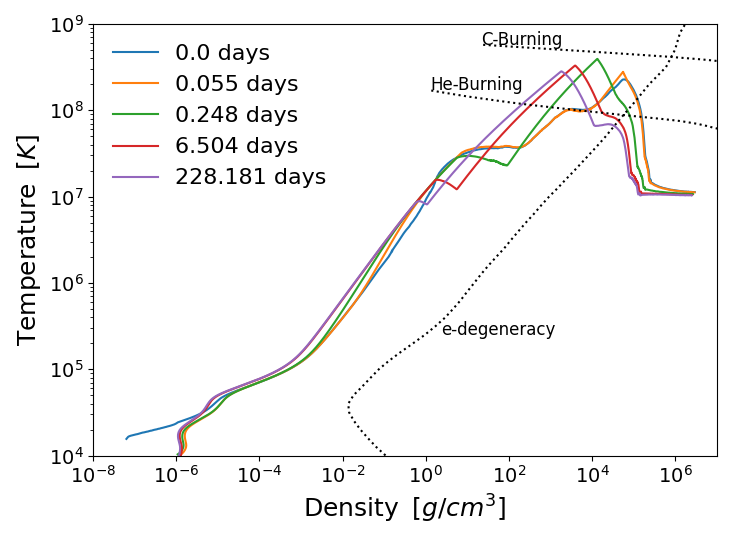}
    \caption{This figure demonstrates the first year of evolution of the Temperature-Density profile.}
    \label{fig:ev_TRho}
\end{figure}

While \cite{Lauer19} do not report surface abundances from model A7, they do report surface \textsuperscript{16}O/\textsuperscript{18}O of about 35 and C/O of about 4.6. This differs from our values of 3.05 and 75.56 significantly. Our \textsuperscript{16}O/\textsuperscript{18}O ratio is in better agreement with observations \citep{Clayton07}, but our C/O is significantly higher than the observed values ($\sim$1). The reason for this is predominantly due to the difference in early evolution. In our models, the temperature profile of the He envelope evolves rapidly during the thermal adjustment phase, causing differences in the early mixing of elements by convection and the peak temperature getting as high as 400 MK for a brief time, which is illustrated in Figure \ref{fig:ev_TRho}. Within the first year of evolution, two distinct convective regions form in the He envelope of the star separated by a temperature inversion. The inner convective region reaches from the He-burning region to the temperature inversion and the outer reaches from the top of the temperature inversion to the surface. This temperature inversion appears to evolve from bumps in the temperature profile which are likely caused by the initial spiral structure of the merger seen in Figure \ref{fig:tempz}. This temperature inversion is present and evolves similarly across all models in this study. We also note that because this first convective gap is eventually bridged in all of our models, its importance to the surface abundance during the RCB phase is minimal. \cite{Lauer19} and \cite{Schwab19} show temperature-density profile plots for engineered models which are much smoother and do not have this temperature inversion present. However, \cite{Schwab19} does not focus on nucleosynthesis and hence mixing is not as important in their study, and \cite{Lauer19} do not perform a detailed analysis of their mixing profile.

\begin{figure}[htb!]
    \centering
    \includegraphics[width=0.99\linewidth]{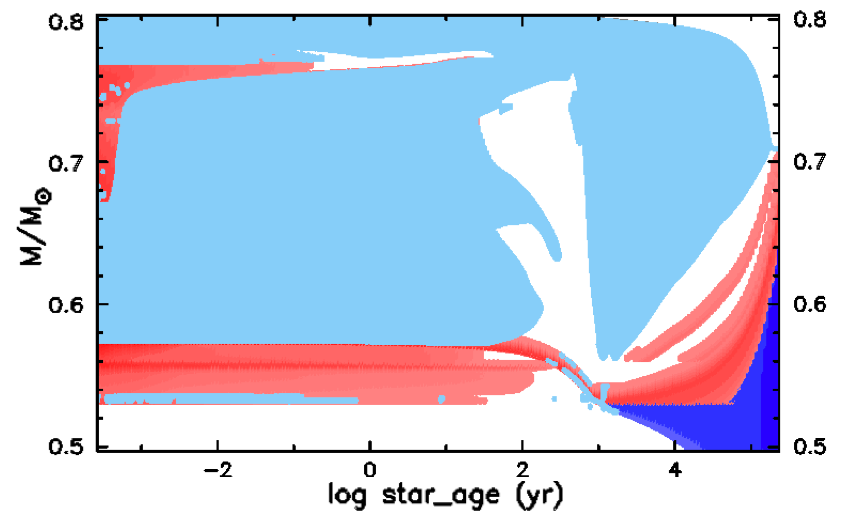}
    \caption{Kippenhahn diagram for Model 1. The vertical axis is the mass coordinate and the horizontal axis is the logarithm of the star age. The blue regions indicate mixing and the red regions indicate nuclear energy generation (darker red being more energy generation). The lighter red region at the top of the envelope and behind the blue region is due to beta decay. The RCB phase occurs at 1600 years.}
    \label{fig:kipp}
\end{figure}

A clearer presentation of how this affects the mixing is shown in Figure \ref{fig:kipp}. The first convective gap caused by the spiral structure can be seen around $0.77 M_\odot$ for the first 40-50 years of evolution. Furthermore, we note that the mixing region between the surface and He-burning shell becomes disconnected after about 30 years, which explains why we see evidence of partial He-burning in RCB stars. The second convective gap is also a feature of the stellar engineered models of \cite{Crawford20}. Because this second convective gap disconnects the surface from the burning region, the surface abundances become set and remain static after that point. This means that the surface abundances are solely dependent on the burning that happens in the first 30 years in our models.

\subsubsection{Robustness of Convective Solution}
Convection is an important feature which brings the partially synthesized material from the burning region to the surface. In order to ensure that convection is a robust feature in our models, we made three test cases with different initial conditions in order to test whether or not convection persists in the envelope. Two of these test cases use a different averaging procedure described in \cite{Endal76}, which averages cells on equipotentials (including the effects of rotation) instead of radial shells. One of the equipotential averaged models uses mass weights during the averaging procedure (replacing $dV_i$ with $dm_i$ in Equation \ref{eq:average}) and the other uses volume weights. The third model uses the density profile obtained by the equipotential averaging procedure with volume weights in order to find a temperature profile in Hydrostatic Equilibrium (HSE). This is done by using the equation for HSE (Equation \ref{eq:HSE}, where $\Psi$ is the effective potential) and the calculated density profile in order to calculate a pressure profile. 

\begin{equation}
\label{eq:HSE}
    \frac{dP}{dr} = -\rho \nabla \Psi
\end{equation}

Then, we use the Helmholtz EoS to calculate temperature given pressure and density and apply the same procedure outlined in Section \ref{sec:methods} in order to insert the degenerate CO core. Figures \ref{fig:r28} and \ref{fig:HSE} show the profiles of Model 1 and the HSE model at key phases during their evolution, respectively.


The purpose of these models is simply to demonstrate that the convective instability persists in the envelope independent of the initial averaging procedure we used. Therefore, it is important to note here that these test models are simplistic in terms of the nuclear network and initial composition. The nucleosynthesis in these models may be unreliable and we therefore choose not to analyze the surface abundances during the RCB phase.

These test cases show that despite starting with a different averaging procedure and therefore a different initial thermal profile, these models converge to similar solutions on the order of 10-100 years. All three test models show the same behavior of two distinct convective regions separated by an initial convective gap that eventually merge and a second convective gap forming around the same time causing only partially synthesized material to mix at the surface of the star. We illustrate this in Figures \ref{fig:r28} and \ref{fig:HSE} for Model 1 and the HSE test model, respectively. In the top panels, we see the initial convective (left) and temperature (right) profiles as well as the initial composition (labeled in the legend). The middle panels show the profiles right after two distinct convective regions form separated by the temperature inversion discussed in Section \ref{sec:surf_and_mix}. The bottoms panels show the moment right after the convective gap closes and synthesized material is allowed to mix to the surface. At this time, the convective region has already receded from the burning region as the star expands and material at the surface will not be synthesized further. This behavior is observed in all of our models regardless of the averaging procedure and demonstrates the robustness of our convective solution. There are, however, fairly large differences differences in the age of the star at which the first convective gap closes (bottom panel) which would result in differences in nucleosynthesis and therefore surface abundances. This is something we aim to explore in more detail in future studies.

\begin{figure}[htb!]
    \centering
    \includegraphics[width=0.99\linewidth]{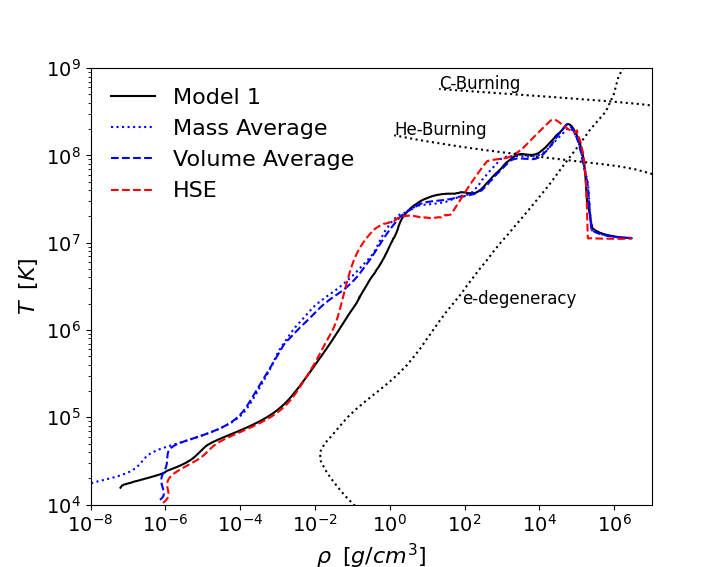}
    \caption{A temperature-density profile comparing the three test cases and Model 1 from Table \ref{tab:1}. The solid black line is Model 1, the blue dotted and dashed lines are mass and volume averaged equipotential models, respectively, and the red dashed line is the model that demands HSE for the averaged density profile.}
    \label{fig:conv_prof}
\end{figure}

\begin{figure*}[htb!]
    \centering
    \includegraphics[width=0.99\linewidth]{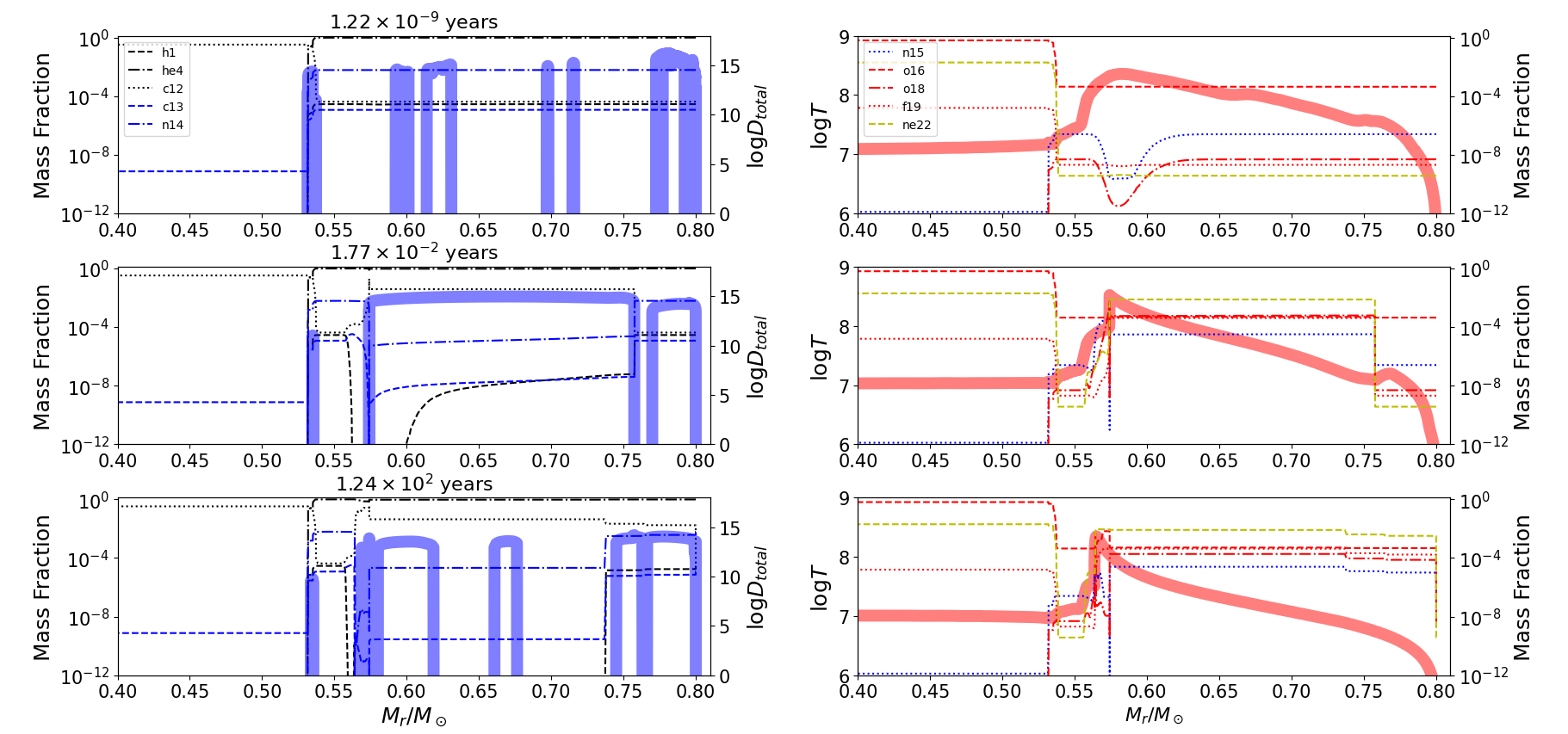}
    \caption{The three panels in this image show the profile of Model 1 at three important phases: the initial relaxed phase, after the two distinct convective regions form, and after the first convective gap closes and the synthesized material is brought to the surface. The surface abundances do not change significantly between that last panel and the RCB phase. The wide red line in the right panels indicates temperature and the wide blue line in the left panels indicates the mixing coefficient.}
    \label{fig:r28}
\end{figure*}
\begin{figure*}[htb!]
    \centering
    \includegraphics[width=0.99\linewidth]{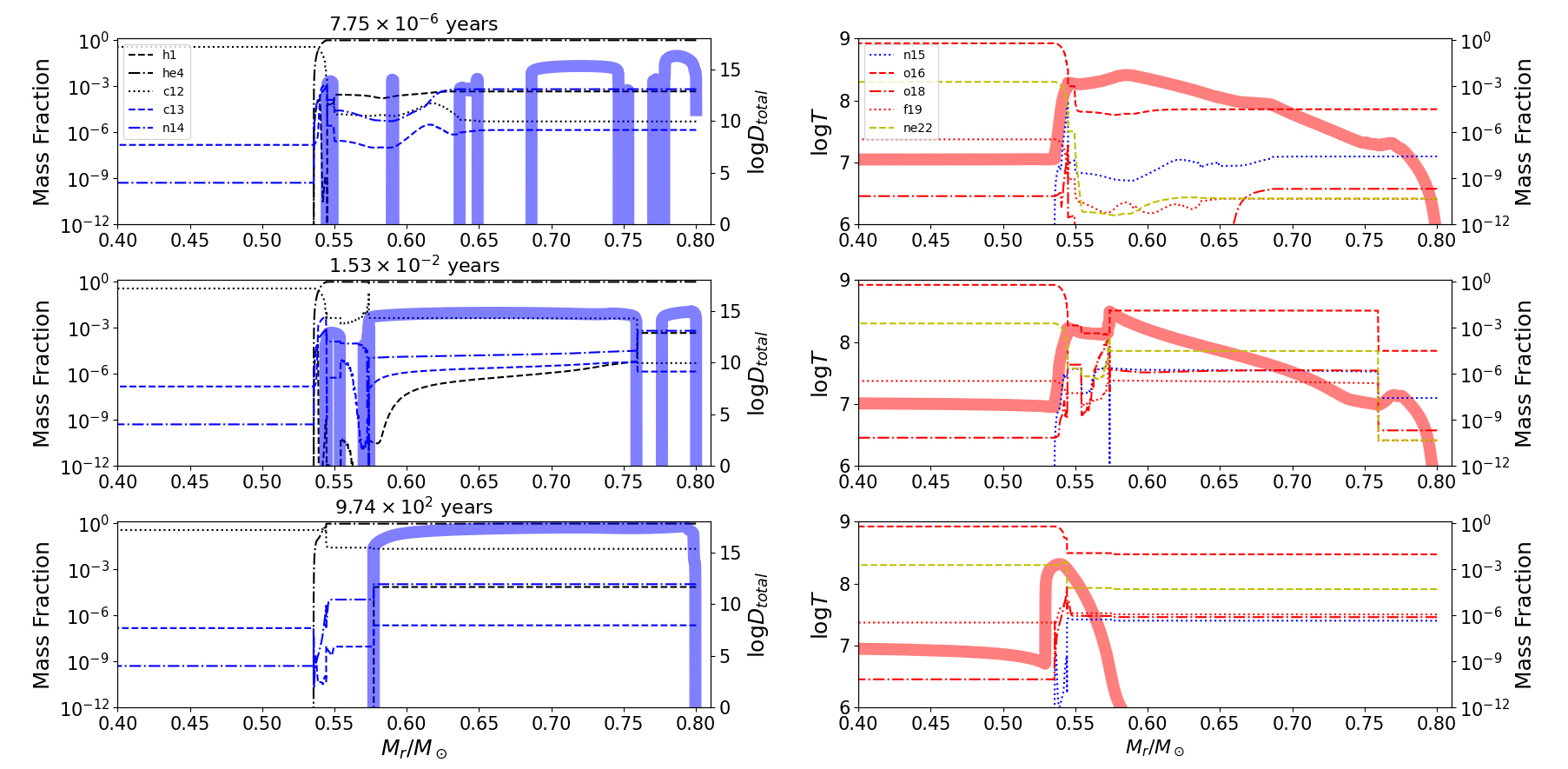}
    \caption{The three panels in this image show the profile of a test model which demands HSE be satisfied for our given density profile. The three phases are described in Figure \ref{fig:r28}, but they occur at different times in this model.}
    \label{fig:HSE}
\end{figure*}

\subsection{Overshooting}
\begin{figure}[htb!]
    \centering
    \includegraphics[width=0.99\linewidth]{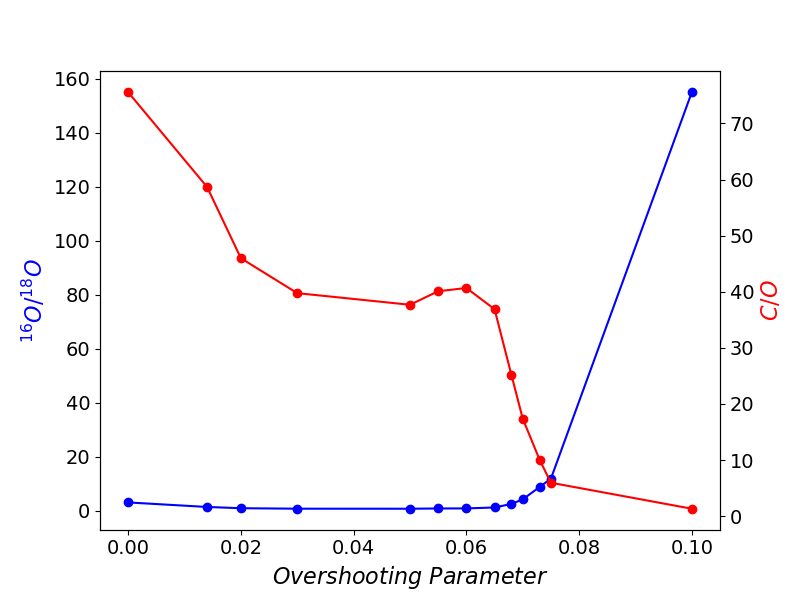}
    \caption{The change in ratios C/O and \textsuperscript{16}O/\textsuperscript{18}O with changing overshooting parameter "\texttt{overshoot\_f}". The model with \texttt{overshoot\_f} of 0.14 is not shown here in order to maintain the scale, but the results can be found in Table \ref{tab:1} Model 20.}
    \label{fig:over_rat}
\end{figure}

\begin{figure*}[htb!]
    \centering
    \includegraphics[width=0.99\linewidth]{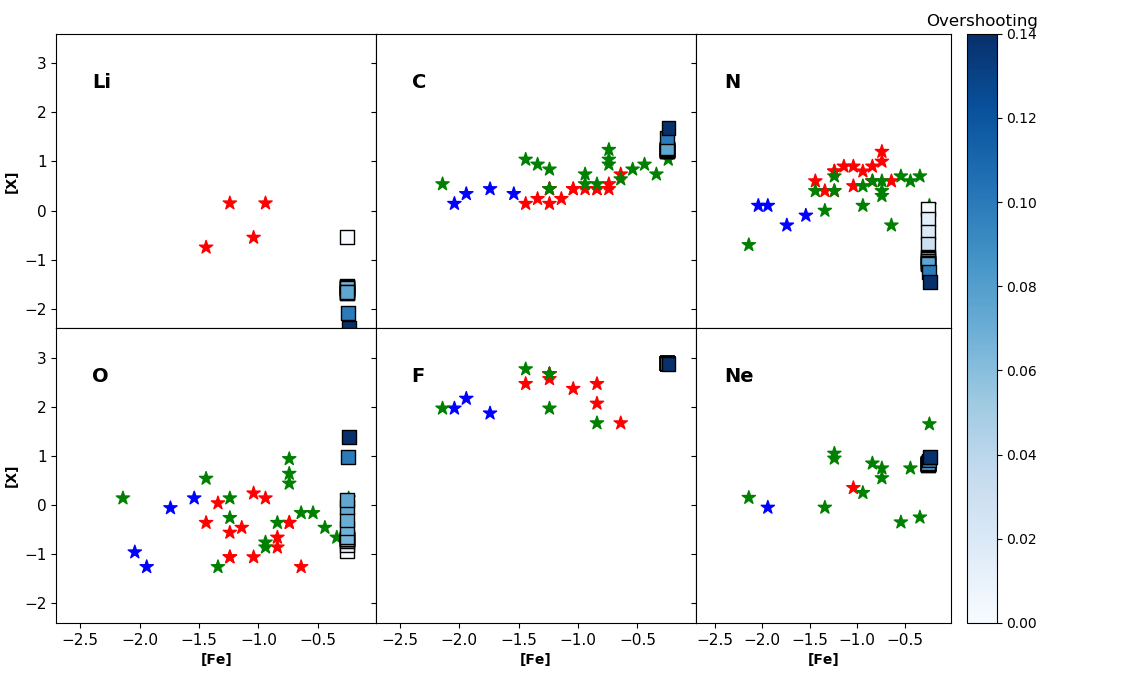}
    \caption{Change in some key surface elemental abundances with changing overshooting parameter "\texttt{overshoot\_f}". The red stars are RCB majority observations (homogeneous in terms of chemical composition), the blue stars are RCB minority observations (diverse in terms of chemical composition), and the green stars are Extreme Helium star (EHe) observations. The units of composition are calculated using Equation \ref{eq:surf}.}
    \label{fig:over_surf}
\end{figure*}

In this study, we vary the strength of overshooting and analyze its effects on the surface composition. \textit{MESA} version r12115 contains many parameters for overshooting which change the depth into the convective region (\texttt{overshoot\_f0}) and the decay length scale outside of the convective region (\texttt{overshoot\_f}) in fractional pressure scale heights. For both overshooting options, there are 12 parameters associated with different regions of the star (non-burning core, non-burning shell (above and below), H-burning core, H-burning shell (above and below), etc). In this study, we vary all \texttt{overshoot\_f} parameters and maintain a constant \texttt{overshoot\_f0} parameter of 0.004. Previous RCB studies have not considered overshooting, but it is an expected physical phenomenon that should be included as one considers more physically motivated models. We maintain overshooting parameters within a reasonable range as discussed in \cite{Stancliffe15}. The results of this study can be found in Figures \ref{fig:over_rat} and \ref{fig:over_surf} as well as Table \ref{tab:1}.

In these results, we see a monotonic decrease in the surface abundance of N with increasing overshooting parameter as seen in Figure \ref{fig:over_surf}. \cite{Crawford20} varied the He-burning shell temperature and observe a decreasing N abundance with an increasing He-burning region temperature. In both cases, \textsuperscript{14}N($\alpha,\gamma$)\textsuperscript{18}F($\beta^+$)\textsuperscript{18}O is being enhanced thus decreasing the amount of N at the surface. This is consistent with the increase in O seen in Figure \ref{fig:over_surf} while \textsuperscript{16}O/\textsuperscript{18}O remains constant up to an overshooting parameter of 0.07, meaning there is an enhancement in the production of \textsuperscript{18}O at the same rate as \textsuperscript{16}O. Beyond an overshooting parameter of 0.07, \textsuperscript{16}O is dramatically enhanced by mixing from the CO core. The \textsuperscript{14}N($\alpha,\gamma$)\textsuperscript{18}F($\beta^+$)\textsuperscript{18}O reaction is paramount in the creation of \textsuperscript{18}O at the levels seen in observations and is discussed in great detail by \cite{Clayton07} and \cite{Menon13}. Lastly, as the overshooting parameter increases, we also see a slight increase in Ne. This happens as \textsuperscript{18}O undergoes $\alpha$-capture and \textsuperscript{22}Ne is created, thus increasing \textsuperscript{16}O/\textsuperscript{18}O even more. \cite{Crawford20} also note these key reactions and show that the sum of N, \textsuperscript{18}O, and Ne is effectively constant as a function of He-burning temperature. This indicates that these isotopes are almost exclusively affected by just the \textsuperscript{14}N($\alpha,\gamma$)\textsuperscript{18}F($\beta^+$)\textsuperscript{18}O($\alpha,\gamma$)\textsuperscript{22}Ne reaction chain.

In this study, we cannot change the He-burning temperature of different models due to the fact that it is self-consistently calculated within the 3D hydrodynamics merger simulation. However, by introducing stronger amounts of overshooting, we are more efficiently bridging the second mixing gap seen in Figure \ref{fig:kipp}. This allows for more time for isotopes to synthesize in the He-burning region and mix to the surface before the convective region breaks from the stellar surface. As the strength of overshooting increases beyond a \texttt{overshoot\_f} value of 0.1, the overshooting beneath the He-burning shell begins to bring up material from the CO core at an overwhelming rate. Since the primary isotopes in the core are \textsuperscript{12}C and \textsuperscript{16}O at a 1:2 ratio, \textsuperscript{16}O/\textsuperscript{18}O increases dramatically while C/O decreases to the order of unity.

The same study is done with sub-solar models for a subset of the overshooting parameters in this study. The results are shown in Table \ref{tab:1}, Models 26-30, and we see similar results in the trend of C/O and \textsuperscript{16}O/\textsuperscript{18}O.


\subsection{Initial Hydrogen abundance}
\label{sec:ini_h1}

\cite{Zhang14} point out the important role of the thin hydrogen envelope in the He WD phase. They point out that if more of this hydrogen envelope were to survive, more \textsuperscript{3}He would also survive. In our models, however, there is also a significant abundance of \textsuperscript{7}Li that survives in the thin hydrogen envelope, which later affects the surface abundances during the RCB phase. 

\begin{figure}[htb!]
    \centering
    \includegraphics[width=0.99\linewidth]{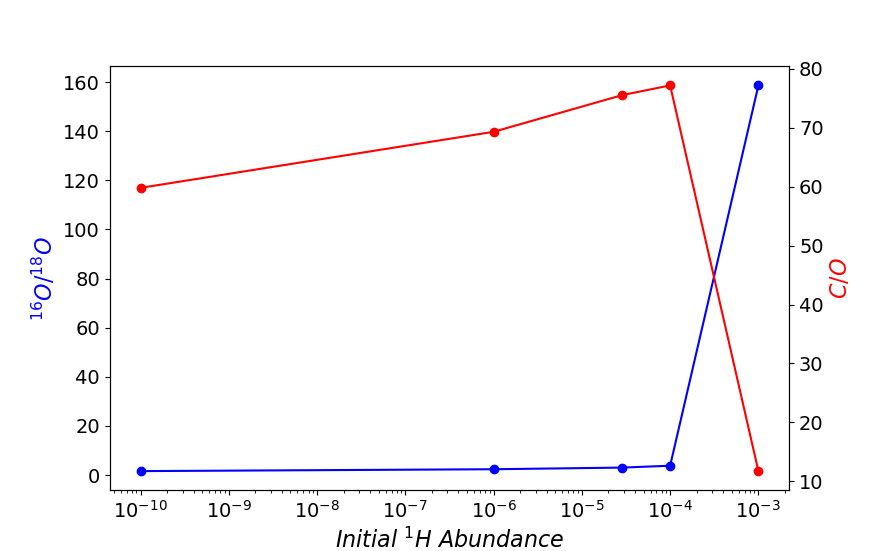}
    \caption{The change in isotopic ratios C/O and \textsuperscript{16}O/\textsuperscript{18}O with changing initial hydrogen abundance in the He envelope. 10\textsuperscript{-99} and 10\textsuperscript{-20} are not shown here to maintain the scale on the x-axis. The results of those models are shown in Table \ref{tab:1}.}
    \label{fig:h_rat}
\end{figure}

\begin{figure*}[htb!]
    \centering
    \includegraphics[width=0.99\linewidth]{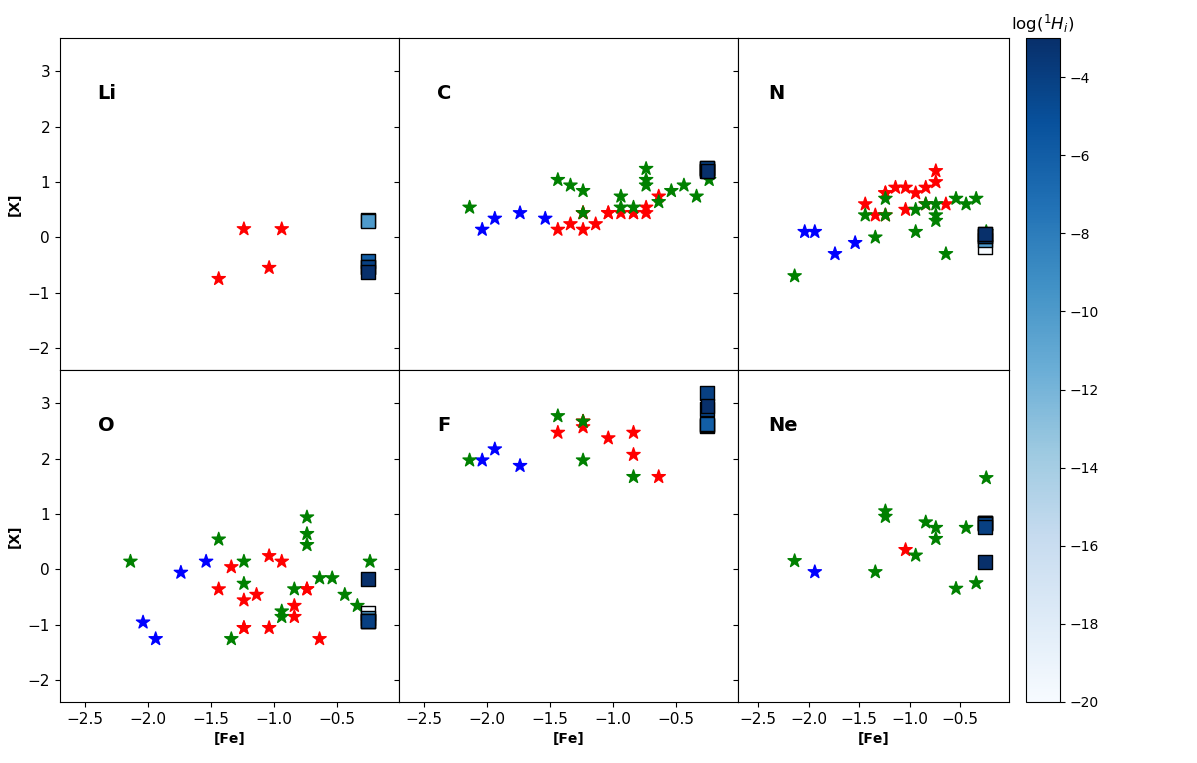}
    \caption{Change in some key surface elemental abundances with changing initial hydrogen abundance in the He envelope. The colored stars are explained in Figure \ref{fig:over_surf}.}
    \label{fig:h_surf}
\end{figure*}

Because we mass average the entire He WD, these abundances in the thin hydrogen envelope have a noticeable effect on the initial uniform abundance of the post-merger He envelope. This is especially true for light elements that are otherwise uncommon in the He WD (\textsuperscript{1}H, \textsuperscript{3}He and \textsuperscript{7}Li, specifically). In order to study the effects of this process, we vary the initial hydrogen abundance in the He envelope while maintaining an \texttt{overshoot\_f} value of 0. The results are shown in Figures \ref{fig:h_rat} and \ref{fig:h_surf}. There are studies such as \cite{Staff12} that show a relationship between He WD mass and the mass of the thin hydrogen envelope and would then set the mass fraction of hydrogen in the envelope of the RCB progenitor. However, we justify varying the initial hydrogen mass fraction because any hydrogen burning during the dynamical merger phase is not included in our simulations since \textit{Octo-tiger} does not include nucleosynthesis. By varying this parameter, we are analyzing the effects of more hydrogen burning during the dynamical phase of the merger assuming the energy generation does not significantly alter the structure of the star.

As expected, the surface abundance of \textsuperscript{7}Li goes down with an increasing hydrogen abundance via \textsuperscript{7}Li($p,\alpha$)\textsuperscript{4}He. Also, there is a large increase in \textsuperscript{16}O accompanied by a significant decrease in \textsuperscript{18}O between the models with initial H mass fractions of 10\textsuperscript{-4} and 10\textsuperscript{-3}. The decrease in \textsuperscript{18}O can be attributed to the enhancement of proton capture reactions on both \textsuperscript{14}N (slightly starving the $\alpha$-capture reaction) and \textsuperscript{18}O. The increase in \textsuperscript{16}O is also due to the enhancement of the proton capture rate on \textsuperscript{14}N. This is clear by the enhancement in \textsuperscript{15}N in the burning region of the higher hydrogen abundance model, which indicates \textsuperscript{14}N($p,\gamma$)\textsuperscript{15}O($\beta^+$)\textsuperscript{15}N is active. From there, the CNO cycle can branch into \textsuperscript{15}N($p,\gamma$)\textsuperscript{16}O or \textsuperscript{15}N($p,\alpha$)\textsuperscript{12}C (which can then $\alpha$-capture or start another CNO cycle). The increase in \textsuperscript{16}O and the decrease in \textsuperscript{18}O explain the sudden decrease in \textsuperscript{22}Ne, seen in Figure \ref{fig:h_surf}, as they starve the \textsuperscript{14}N($\alpha,\gamma$)\textsuperscript{18}F($\beta^+$)\textsuperscript{18}O($\alpha,\gamma$)\textsuperscript{22}Ne reaction chain of the initial \textsuperscript{14}N $\alpha$-capture.


\subsection{NuGrid Post-Processing}
\begin{figure*}[htb!]
    \centering
    \includegraphics[width=0.99\linewidth]{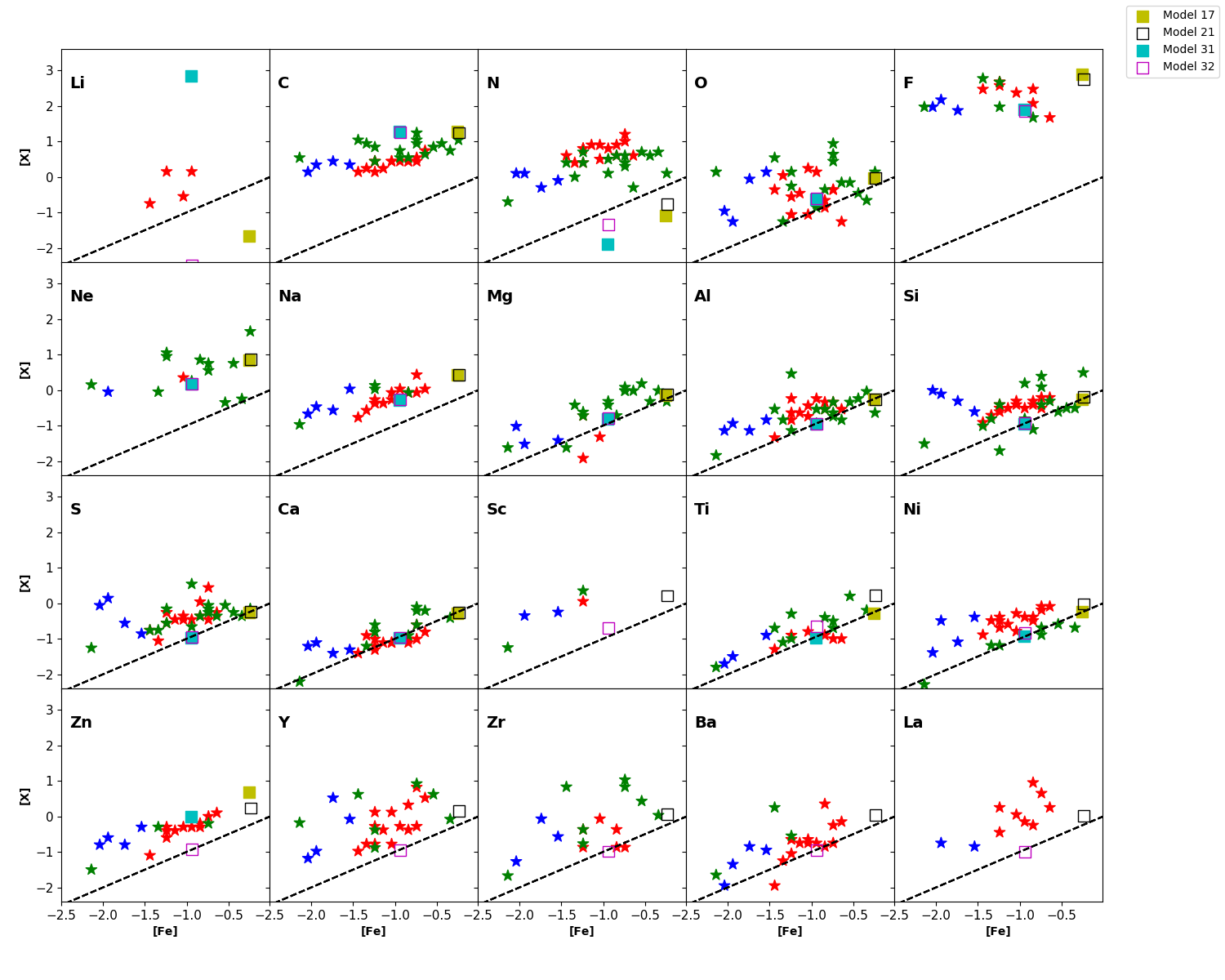}
    \caption{Surface Abundances calculated by \textit{MESA} (filled in lime and cyan) compared with the result post-processed by \textit{MPPNP} (empty black and magenta). The model numbers in the legend refer to the models presented in Table \ref{tab:1}. The dashed black line represents scaled solar metallicities. The colored stars are explained in Figure \ref{fig:over_surf}.}
    \label{fig:surf_abund}
\end{figure*}

The final section of this study discusses post-processed nucleosynthesis on a much larger nuclear network containing 1093 isotopes and over 14,000 reactions. We chose two models, one with solar and one with sub-solar metallicity, that are most consistent with previous stellar engineering attempts and observations (Models 17 and 31, respectively). Figure \ref{fig:surf_abund} shows the results of these models and the \textit{MPPNP} post-processed results. Below we explain some of the differences between the \textit{MESA} and \textit{MPPNP} models, including surface Li, N, and s-process elements.

\subsubsection{Lithium}
\label{sec:nu_li}
The biggest difference between \textit{MPPNP} and \textit{MESA} surface abundances in Figure \ref{fig:surf_abund} is the surface abundance of \textsuperscript{7}Li. This is observed in both the solar and sub-solar models. \textit{MPPNP} burns about 6 orders of magnitude more \textsuperscript{7}Li because it includes the \textsuperscript{7}Li($\alpha,\gamma$)\textsuperscript{11}B reaction while \texttt{mesa\_75} does not. Typically, \textsuperscript{7}Li would be burned immediately by proton capture reactions, but in both the solar and sub-solar models, there is very little initial hydrogen in the He envelope. This low amount of initial hydrogen causes other problems, specifically, in the sub-solar case where the initial hydrogen is 10\textsuperscript{-10} which is further discussed in the Section \ref{sec:sproc}.

It is important to note that in \textit{MESA} version r12115, the default \textsuperscript{7}Li($p,\alpha$)\textsuperscript{4}He reaction has a sudden cutoff at a temperature of 10 MK. When simulating the He WD progenitor, this cutoff results in a fairly large amount of \textsuperscript{7}Li in the hydrogen envelope which produces an unusually high initial abundance in the He envelope of the post-merger. In reality, the \textsuperscript{7}Li should have likely been burned away during the evolution of the He WD or during the tidal disruption phase of the merger event. However, since \textit{Octo-tiger} does not perform nucleosynthesis or trace the species from each progenitor, the nuclear burning in the merger phase is not considered.

Given that the post-merger object in \textit{MESA} is not burning \textsuperscript{7}Li via $\alpha$-capture and the initial \textsuperscript{7}Li abundance should be much lower for the above reasons, we do not expect to see any measurable amount of \textsuperscript{7}Li on the surface of our models. This is demonstrated by the results of the \textit{MPPNP} post-processing and is in agreement with all but four RCB observations \citep{Jeffery11}. It is still difficult to explain why \textsuperscript{7}Li is abundant in these four observations given the expectation that it all be burned during the merger process \citep{Clayton07}. We do not explore the lithium problem any further as it is outside the scope of this paper.

\vspace{1cm}
\subsubsection{Nitrogen}
The surface N abundances in \textit{MPPNP} are consistently higher than in \textit{MESA} and we attribute this to the burning of \textsuperscript{7}Li via $\alpha$-capture. \textsuperscript{14}N can be created in \textit{MPPNP} via the \textsuperscript{7}Li($\alpha,\gamma$)\textsuperscript{11}B($\alpha,n$)\textsuperscript{14}N reaction chain, while the \texttt{mesa\_75} network does not contain \textsuperscript{11}B. This is consistent with the fact that we see a larger enhancement of \textsuperscript{14}N in the sub-solar case where there is more \textsuperscript{7}Li to be burned.

It is worth mentioning that although all four models have lower surface N abundances than observations, this result is still consistent with the engineered models of \cite{Crawford20}. Their hot models ($>$320 MK) also show diminished \textsuperscript{14}N and attribute that to enhanced $\alpha$-capture. Our models reach a maximum temperature around 400 MK during the thermal adjustment of the envelope which coincides with the time where most of the nucleosynthesis takes place. This decrease in \textsuperscript{14}N is also associated with a sudden burst of neutrons from the \textsuperscript{22}Ne source, which becomes active above ~250 MK \citep{Kappeler11}. Following the neutron burst, \textsuperscript{14}N then acts as a neutron poison and is rapidly burned by neutron capture reactions. Both enhancements in $\alpha$-capture and n-capture reactions would then result in the observed enhancements of \textsuperscript{18}O and \textsuperscript{19}F, respectively. \textsuperscript{14}N can be replenished to observed levels by adding a higher initial hydrogen abundance to the He envelope, but as seen in Figure \ref{fig:h_rat}, this increases the oxygen isotopic ratio beyond the acceptable range.

\subsubsection{s-process Elements}
\label{sec:sproc}
One characteristic of RCB stars is their enhancement in s-process elements on the surface \citep{Jeffery11}. Previous studies in stellar engineering and our \textit{MESA} models do not include s-process nucleosynthesis, but \textit{MPPNP} does. Figure \ref{fig:surf_abund} shows that the solar case has a slight enhancement of s-process elements (Sc, Y, Zr, Ba, and La), but the sub-solar case does not see this enhancement. The reason we are not seeing s-process enhancement is because in both cases we have very little \textsuperscript{13}C neutron source, which would be active above a temperature of 100 MK \citep{Kappeler11}. This is a direct result of having a lower initial hydrogen abundance as the \textsuperscript{12}C being brought up from the core cannot proton capture to create \textsuperscript{13}C via \textsuperscript{12}C($p,\gamma$)\textsuperscript{13}N($\beta^+$)\textsuperscript{13}C. With a higher hydrogen abundance, we would expect to see more enhancement of s-process elements, but would also push the oxygen ratio outside the observed range (see Figure \ref{fig:h_rat}).

\section{Conclusions}
\label{sec:conclusion}
This study presents the first attempt to reproduce RCB star surface abundances (including s-process elements) starting from a 3D hydrodynamics merger simulation. While this attempt is not complete and negates some potentially important phases of nuclear burning (namely, during the merger process), it establishes a mechanism for bringing models from a 3D merger simulation to a full scale 1D nucleosynthesis post-processing network. Our models show strong similarities to the stellar engineering models of \cite{Crawford20} and \cite{Lauer19} in terms of evolution, mixing, and most surface abundances. Our models also have difficulties in matching \textsuperscript{14}N to observations (similar to \cite{Crawford20}) and generating s-process elements.

Our early stage evolution strongly resembles those of \cite{Lauer19}, \cite{Schwab19}, and \cite{Crawford20}. This is because the stellar engineering process is informed by 3D hydrodynamics models, but does not evolve directly from the results of those models. The differences between those models and the He star model of \cite{Weiss87} and \cite{Menon13, Menon19} are mostly due to how the initial state of the envelope reacts to the input luminosity of the He-burning shell.

While overshooting was necessary in order to match the isotopic ratios, it is a realistic physical phenomenon expected to occur in stars where convection operates and should be included by default in stellar models. Overshooting was not previously studied in the context of RCB stars, but \cite{Stancliffe15} use models with parameters in the same range as our overshoot parameters. The parameter space of overshoot parameters could be further constrained by 3D models with realistic mixing procedures, but the adopted MLT prescription in \textit{MESA} is currently the best we can achieve. We find that an overshooting parameter of 0.073 and 0.068 for the solar and sub-solar metallicity models, respectively, yield reasonable agreement to observations and previous studies in surface composition. Lower values tend to increase C/O outside the acceptable range while higher values tend to increase \textsuperscript{18}O/\textsuperscript{16}O and decrease surface N outside the acceptable range. There are, however, many other parameters to study that will change the surface composition.

This study also explores the effects of changing the initial hydrogen abundance. The initial hydrogen abundance may change based on how much of the hydrogen shell of the He WD progenitor survives during the merger process. While studies such as \cite{Staff12} or \cite{Driebe98} constrain the mass of the hydrogen envelope to values much higher than our models, we note that \textit{Octo-tiger} does not perform nucleosynthesis during the dynamical merger phase, which may burn a significant fraction of the initial hydrogen. Of course, with too little initial hydrogen (our sub-solar model, for instance) the He-burning shell cannot create a sustainable \textsuperscript{13}C neutron source. Additionally, there are common envelope phases before the WD merger that are not simulated in this study but will affect the progenitor composition. The complexities of the complete RCB evolution necessitates a wider parameter space study of the initial post-merger composition. Future studies which pursue the 3D simulation of the merger being mapped into 1D nuclear lifetime evolutionary codes like \textit{MESA} should include a basic nuclear network that accounts for most pp chain and CNO cycle elements during the merger phase. Being able to trace and burn isotopes during the merger will give improved compositional profiles to be mapped into a 1D evolution code.

In future work, we plan to continue to use this approach to produce more realistic models of RCB stars. Improvements we plan to make include using a basic nuclear network during the merger phase, including a prescription for angular momentum diffusion in order to naturally spherize the merger, creating a customized RCB nuclear network which includes important species of this study for speedup of the \textit{MESA} simulations (i.e. we do not need to co-process elements beyond \textsuperscript{26}Mg at our burning temperatures), and a more careful consideration of opacities and mass loss during the RCB evolution. This is the first in a long line of improvements to be made in modeling RCB stars more realistically. Our attempts to produce better models also necessitates the inclusion of more detailed physical processes during the merger phase, which we will attempt to address in future studies.

\acknowledgements{We would like to acknowledge \textit{MESA} community for quick troubleshooting help and advice. We thank Josiah Schwab and Pablo Marchant for useful discussion. We also thank the reviewer for constructive advice and insight. E.C. would like to thank the National Science Foundation for its support through award number AST-1907617 and the Louisiana State University College of Science and the Department of Physics \& Astronomy for their support. G.C. thanks the NSF for its support through award 1814967. This research has used the Astrohub online virtual research environment (https://astrohub.uvic.ca).}

\software{Octo-tiger \citep{Marcello16}, MESA-v12115 \citep{Paxton2011,Paxton2013,Paxton2015,Paxton2018,Paxton2019}, MPPNP \citep{Herwig08}, Python (available from python.org), Matplotlib \citep{matplotlib}, Numpy \citep{numpy}}

\bibliography{main.bib}
\end{document}